\documentclass[aps,showpacs,groupedaddress,prb,twocolumn]{revtex4-1}

\usepackage{graphicx}
\usepackage{dcolumn}
\usepackage{bm}
\usepackage[T1]{fontenc}
\usepackage{color}
\usepackage[dvips, colorlinks=true, bookmarks, pdftitle={}, pdfauthor={}, pdfsubject={}, pdfkeywords={PDF, LaTeX, hyperlinks, hyperref}]{hyperref}

\newcommand{\trans}{{\cal T}}
\newcommand{\diff}{{\cal D}}

\begin{document}
\title{Phonon Transport in Large Scale Carbon-Based Disordered Materials:\\ 
Implementation of an Efficient Order-\textit{N} and Real Space Kubo Methodology}
\author{Wu Li$^{1,2}$}
\author{Haldun Sevin\c{c}li$^{2}$}
\author{Gianaurelio Cuniberti$^{2}$}
\author{Stephan Roche$^{2,3,4}$}
\affiliation{$^1$Institute of Physics, Chinese Academy of Sciences, 100190 Beijing, China.}
\affiliation{$^2$Institute for Materials Science and Max Bergmann Center of Biomaterials, Dresden 
University of Technology, 01062 Dresden, Germany.}
\affiliation{$^3$CIN2 (CSIC-ICN), 
08193 Bellaterra, Barcelona, Spain.}
\affiliation{$^4$CEA, INAC, SP2M, L\_Sim, 17 avenue des Martyrs, 38054 Grenoble, France}

\begin{abstract}
We have developed an efficient order-$N$ real space Kubo approach for the calculation of the phonon conductivity which outperforms state-of-the-art alternative implementations based on the Green's function formalism.
The method treats efficiently the time-dependent propagation of phonon wave packets in real space, and this dynamics is related to the calculation of the thermal conductance. 
Without loss of generality, we validate the accuracy of the method by
 comparing the calculated phonon mean free paths in disordered carbon nanotubes 
(isotope impurities) with other approaches, and further illustrate its upscalability by exploring the thermal conductance features in large width edge-disordered graphene nanoribbons (up to $\sim 20$nm), which is out of the reach of more conventional techniques. 
We show that edge-disorder is the most important scattering mechanism for phonons in graphene nanoribbons with realistic sizes and thermal conductance can be reduced by a factor of $\sim$10.
\end{abstract}

\pacs{72.80.Vp,72.15.Rn,73.22.Pr}

\maketitle

\section{Introduction}
In recent years, the understanding of phonon transport in carbon-based materials such as carbon nanotubes (CNTs) \cite{RMP} and graphene-based materials \cite{Graphene} has become particularly important, both for fundamental studies of coherent transport and also in view of novel applications. 
The thermal conductivity of suspended and supported single graphene layers has been found to be extremely high \cite{balandin,RuoffScience} owing to micrometer long phonon mean free paths (MFP).
Such high thermal conductivity of two-dimensional graphene increases its potential for faster nano-electronic devices with less energy dissipation \cite{pop}.
On the other hand, damaging a material like graphene could present interesting opportunities like achieving a high thermoelectric figure of merit and observation of Anderson localization of phonons.
The question whether or not Anderson localization of acoustic phonons can be demonstrated unambiguously in disordered materials has been a long-standing problem in phonon physics,
a phenomenon originating from the interference between multiple scattering paths was found ubiquitous in wave physics \cite{Anderson,Scheffold99}.
Besides, disordered CNT-based bundles have been found to exhibit a tendency towards a thermal insulating regime \cite{cagin,MingoCNTAL}.
Also, isotope disorder was shown to strongly impact on the high-energy phonon modes, resulting in very low mean free paths, in marked contrast with the genuine robustness of ballistic conduction  for low-energy phonon modes \cite{Savic}. 
As a result, the contribution of quantum localization effects of high energy modes was found to be important but completely masked when considering the thermal conductance of the disordered material (being an integrated quantity over the entire phonon spectrum).
Graphene nanoribbons (GNRs) offer an alternative to carbon nanotubes. 
By constructing heterostructures from pristine graphene mixed with disordered (isotope impurities) GNRs \cite{MingoPRBh}, or selective coverage of hydrogen impurities \cite{baowenli}, it is possible to tune  the resulting phonon transport capability. 
Another important source of disorder is provided by unavoidable ribbon edge irregularities, absent in CNTs which have already been shown to significantly reduce thermal conductance in small width GNRs \cite{Haldun,Lan}.

Exploring such possibilities, however, demand the development of efficient computational approaches which are able to tackle large scale (and realistic) simulations of material of interest.
The Green's function (GF) methods are able to incorporate microscopic details of the system, but they require matrix inversions which unavoidably limit the accessible size of the simulated systems. 
For the case of GNRs, although the system length can be upscaled without difficulty thanks to the decimation procedure, the computational cost becomes prohibitive for lateral sizes above 10 nm. \cite{Savic,Haldun}

In this Rapid Communication, using the real space Kubo (RSK) formalism we first demonstrate that a time-dependent phonon wave packet formalism can be connected to the calculation of the thermal conductance.
After validating this numerical approach by comparing the obtained phonon MFPs in disordered CNTs (with isotope impurities) with previously computed ones by means of GF-based method \cite{Savic}, we apply the new algorithm to large width GNRs, and focus on the impact of edge-disorder profiles.
Scaling properties of phonon MFP and temperature-dependent thermal conductance are calculated as a function of edge-disorder strength and for lateral ribbon sizes accessible to today's state-of-the-art lithography. 
The broad generality of this method could offer a novel framework to explore other types of complex materials.

\begin{figure}[t]
\begin{center}\leavevmode
\includegraphics[scale=1]{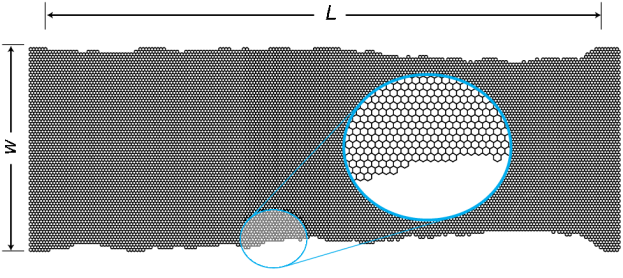}
\caption{(Color online) A short portion of an edge-disordered ZGNR  with width $w$=17.04 nm, length $L\sim50$~nm and disorder density of 10\%. In the calculation, the length of ZGNR is chosen to be 983.80nm, and periodic boundary condition is employed.}
\label{fig1}
\end{center}
\end{figure}

\section{Computational phonon transport methodology}
To investigate bulk quantum phonon transport in disordered materials, the use of the Kubo formalism turns out to be the most natural and computationally efficient one. 
It has already been used for investigating thermal transport in disordered binary alloys or nanocrystralline silicon \cite{Kubophonon,Allen}. 
Similarly, many efforts have been made to explore time-dependent features of propagating phonon (or polaron) wave packets in complex quantum systems \cite{WPP}. 
Here a novel real space implementation of the Kubo formula for phonon propagation is given, establishing a direct computational bridge between phonon dynamics and the thermal conductance. 
In comparison to the previous methods, we extract the dynamical information from time evolution of the wave packet \cite{Allen} and simulate Hamiltonian dynamics in place of Newtonian equations of motion \cite{loh}, so that only one initial condition is required, i.e. the initial atomic displacements, and one does not need to calculate the atomic velocities in time.
Additionally, by using the Lanczos technique which avoids any matrix inversion, a considerable computational efficiency gain is obtained, allowing the study of very large scale materials. 
The starting Hamiltonian which describes the phonon spectrum, taking only the harmonic interactions into account, is of the form 
\begin{equation} \label{Hamiltonian}
\mathcal{H}=\sum_{i} \frac{\hat{p}_{i}^{2}}{2M_{i}}+\sum_{ij} \Phi_{ij}\hat{u}_{i}\hat{u}_{j},
\end{equation}
where $\hat{u}_{i}$ and $\hat{p}_{i}$ are the displacement and momentum operators for the $i$th degree of freedom, $M_{i}$ is the corresponding mass, and $\Phi$ is the force constant tensor. 
Based on the linear response theory, the phonon conductivity $\sigma$ can be defined as
$\Omega T^{-1}\int_{0}^{\beta}d\lambda\int_{0}^{\infty}dt\langle \hat{J}^{x}(-i\hbar\lambda)\hat{J}^{x}(t)\rangle$ \cite{Allen}. 
$\hat{J}^{x}$ is the $x$ component of the energy flux operator $\hat{\textbf{J}}$, and it can be expressed as $\hat{J}^{x}={1}/{2\Omega}\sum_{ij} (X_{i}-X_{j})\Phi_{ij}\hat{u}_{i}\hat{v}_{j}$, where $\hat{v}_{j}$ is the velocity operator and $X_{i}$ is the equilibrium position of the atom to which the $i$th degree of freedom belongs. 
After some algebra, $\sigma$ becomes
\begin{equation}
\sigma=-\frac{\pi}{\Omega}\int\limits_{0}^{\infty}
\mathrm{d}\omega\,\hbar\omega
\frac{\partial f_B}{\partial T}
\textrm{Tr}\{[\hat{X},D]\delta(\omega^{2}-D)[\hat{X},D]\delta(\omega^{2}-D)\},
\end{equation}
where $f_B$ is the Bose distribution function, and $D_{ij}=\Phi_{ij}/{\sqrt{M_{i}M_{j}}}$ is the mass normalized dynamical matrix, $\mathrm{Tr}$ denotes the trace of the matrix, and $\delta(\omega2-D)$ is the Dirac-delta operator. .
The dynamical matrix can be obtained given the atomistic configuration of the system and it defines a Newtonian equation of motion which is of second order in time.
Using the fact that the operators $D$ and $H$ have the same energy spectrum ($\hbar^2D=H^2$) and defining $\hat{V}_x=[\hat{X},H]/i\hbar$, one can write the thermal conductance of a one-dimensional system as
\begin{equation}
\kappa=\frac{\pi\hbar^2}{L^{2}}
\int_{0}^{\infty}\mathrm{d}\omega\,\hbar\omega
\frac{\partial f_B}{\partial T}
\textrm{Tr}\{\hat{V}_{x}\delta(\hbar\omega-H)\hat{V}_{x}\delta(\hbar\omega-H)\}.
\end{equation}
The thermal conductance can also be derived from the Landauer formalism \cite{Rego_Mingo} or the nonequilibrium Green's function approach \cite{ciraci_yamamoto} as
\begin{equation}
\kappa=
\frac{1}{2\pi}
\int_{0}^{\infty}\mathrm{d}\omega\,\hbar\omega
\frac{\partial f_B}{\partial T}
\trans(\omega),
\label{kappa}
\end{equation}
with $\trans(\omega)$ being the phonon transmission function.
Comparing the two formulas, the transmission function is defined as
\begin{equation}\label{KuboTrans}
\trans(\omega)=\frac{2\pi^2\hbar^2}{L^2}\textrm{Tr}\{\hat{V}_{x}\delta(\hbar\omega-H)\hat{V}_{x}\delta(\hbar\omega-H)\},
\end{equation}
which has the same form as the electron transmission function derived from the Kubo-Greenwood formula \cite{stephan1},
\begin{equation}\label{KuboTransEl}
\trans_{el}(E)=\frac{2\pi^2\hbar^{2}}{L^{2}}\textrm{Tr}\{\hat{V}_{x}\delta(E-H_{el})\hat{V}_{x}\delta(E-H_{el})\}.
\end{equation} 
This equivalence allows us to implement an order $N$ algorithm related to the Lanczos method, which has been successfully applied to electron conduction in complex materials \cite{stephan1}. 
The Lanczos method has been previously employed in the past for calculating the vibrational density of states of disordered systems \cite{DoSv}.
The starting point of the RSK method is the transmission function as expressed in Eq.~(\ref{KuboTransEl}). 
One can rewrite $\trans(\omega)$ in terms of the diffusion coefficient,
\begin{equation}
\trans(\omega)=
\frac{2\omega\pi}{L^2}
\mathrm{Tr}\left\{\delta(\omega^2-D)\right\}
\lim_{t\to\infty}\frac{d}{\mathrm{d}t}
(t\mathcal{D}(\omega,t)),
\end{equation}
where the diffusion coefficient $\mathcal{D}$ has the form
\begin{equation}\label{diffusion}
\mathcal{D}(\omega,t)=\frac{1}{t}\frac{\mathrm{Tr}\left\{(\hat{X}(t)-\hat{X}(0))^2\delta(\omega^2-D)\right\}}
{\mathrm{Tr}\left\{\delta(\omega^2-D)\right\}},
\end{equation}
where $\hat{X}(t)$ is the position operator in the Heisenberg picture.
In the diffusive regime
$\mathcal{D}(\omega,t)=\mathcal{D}_\mathrm{max}(\omega)$, so that the
transmission function reduces to
\begin{equation}
\trans(\omega)=\frac{2\omega\pi}{L^2}\mathrm{Tr}\left\{\delta(\omega^2-D)\right\}\mathcal{D}_\mathrm{max}(\omega),
\label{eqn:trans:fg}
\end{equation}
where in the ballistic regime $\mathcal{D}(\omega,t)=v^2(\omega)t$, with
$v(\omega)$ being the average group
 velocity over states with frequency $\omega$. Since the number of
channels is
$N_\mathrm{ch}(\omega)\approx
2\omega\pi\mathrm{Tr}\left\{\delta(\omega^2-D)\right\}v(\omega)/L$,
the phonon mean free path can be approximated by $\ell(\omega)=\mathcal{D}_\mathrm{max}(\omega)/v(\omega)$. 
By computing $\mathcal{D}(\omega,t)$, one can thus deduce $\mathcal{D}_\mathrm{max}(\omega)$ and $v(\omega)$
and $\ell(\omega)$. 
Note that the numerator in Eq.~(\ref{diffusion}) can be rewritten as 
$\mathrm{Tr}\{[X,U(t)]^{\dag}\delta(\omega^2-D)[X,U(t)]\}$, 
and approximated by 
$\mathrm{Tr}\{[X,\mathcal{U}(\tau)]^{\dag}\delta(\omega^2-D)[X,\mathcal{U}(\tau)]\}$ to the first order in perturbation theory, 
with $\mathcal{U}(\tau)=\mathrm{e}^{-iD\tau}$, and $\tau=t/2\omega$.
The trace can be efficiently calculated through an average over a few random phase states of atomic displacements as
$N\overline{\langle\psi|[X,\mathcal{U}(\tau)]^{\dag}\delta(\omega^2-D)[X,\mathcal{U}(\tau)]|\psi\rangle}$,
$N$ being the number of degrees of freedom and the bra-ket corresponds to the local density of states (LDOS) associated with the vector $[X,\mathcal{U}(\tau)]|\psi\rangle$
which is calculated by  using Chebyshev expansion of $\mathcal{U}(\tau)$.
LDOS is obtained by using the Lanczos method,
and $\mathrm{Tr}\left\{\delta(\omega^2-D)\right\}$ is also calculated by averaging the LDOS of $|\psi\rangle$ through the Lanczos method.

\begin{figure}[t]
\begin{center}\leavevmode
\includegraphics[scale=1]{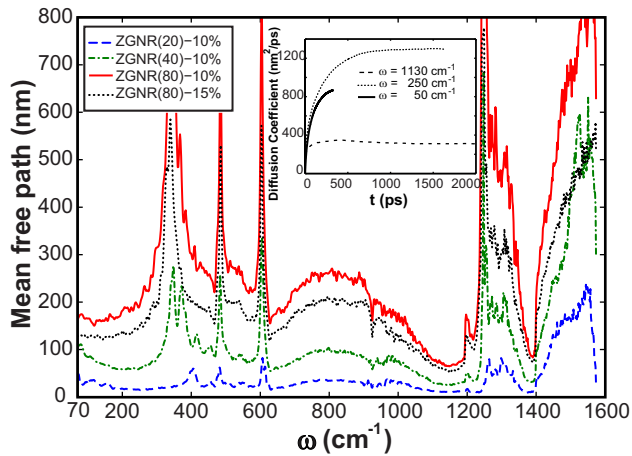}
\caption{(Color online) 
Elastic MFP for ZGNR of widths $N_z=$20, 40, and 80 (4.26, 8.52, and 17.04 nm, respectively) with disorder density of 10\%, and also for the $N_z=$80 and 15\% disorder for comparison.
Inset: Time dependent diffusion coefficients $\diff(\omega,t)$ for three chosen frequencies. The same calculation time corresponds to different evolution time for different frequencies.}
\label{fig2}
\end{center}
\end{figure}

\section{Results}
As a test case, we first consider a CNT(7,0) with 10.7\% $^{14}$C impurities, which was studied in Ref. \onlinecite{Savic} using the GF method.
From the saturation values of the time-dependent diffusion coefficients, we obtain the phonon MFP using the RSK scheme which compares very well with those obtained from the GF method \cite{haldun_tbp}.

\begin{figure}[t]
\begin{center}\leavevmode
\includegraphics[scale=1]{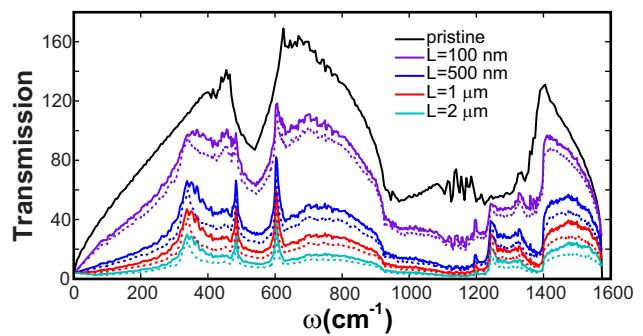}
\caption{(Color online) Transmission spectra for different lengths of ZGNR(80) with edge-disorder of 10\% (solid) and 15\% (dashed).}
\label{fig3}
\end{center}
\end{figure}

\begin{figure}[t]
\begin{center}\leavevmode
\includegraphics[scale=1]{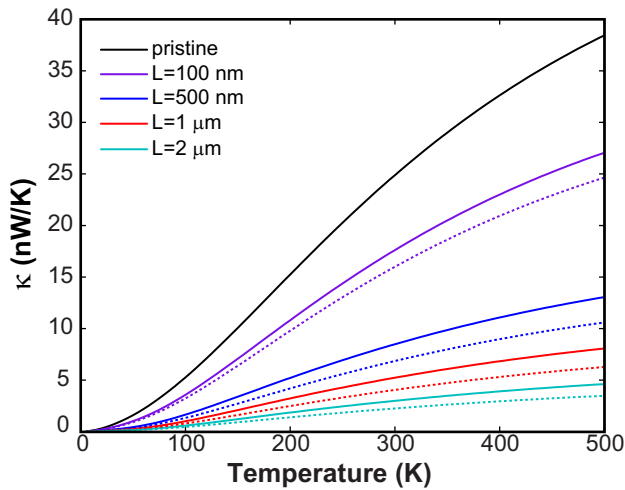}
\caption{(Color online) Temperature and length-dependent behaviors of the ZGNR(80) thermal conductance (10\% disorder (solid lines), 15\% (dashed lines).}
\label{fig4}
\end{center}
\end{figure}

Next, we consider zigzag graphene nanoribbons (ZGNR) of different widths with edge-disorder. 
We use the fourth nearest neighbor force constants for building the dynamical matrices \cite{saito_zimmermann}.
The ribbon widths are defined with the number of zigzag chains $N_z=20$, 40, and 80, and the relative amount of edge defects (removed carbon atoms at the edges) is chosen to be  10\%, and additionally 15\% for $N_z=80$ (Fig.~\ref{fig1}).
The inset of Fig.~\ref{fig2} displays the evolution of the wave packet dynamics for different frequencies for ZGNR(80) with 10\% edge-disorder..
The linear increase in $\diff(\omega,t)$ at $t>0$ indicates ballistic transport at relatively short distances, whereas the decrease in $\diff(\omega,t)$ is a signature of localization for this particular frequency.
The fact that $\diff$ decreases slowly suggests the localization length is large.
The saturation of $\diff(\omega,t)$ to a maximum value characterizes diffusive transport (c.f. Eq.~\ref{eqn:trans:fg}).
In Fig.~\ref{fig2}, the decay of $\ell(\omega)$ with decreasing the ribbon width is shown at a fixed disorder strength.
This behavior can be rationalized with the fact that the scattering rate decreases with increasing width, a behavior previously derived for electron transport in both disordered CNTs and GNRs \cite{TNT}.
One notes that, for low frequency modes, the MFP are several hundreds of nanometers, and due to large values of $N_{\mathrm{ch}}$ the possibility to observe any onset of Anderson localization is jeopardized in the thermal conductance, as previously discussed for small diameter disordered carbon nanotubes \cite{Savic}. 
We then focus on ZGNR(80), and obtain the transmission according to $\mathcal{T}(\omega)=N_\mathrm{ch}/(1+L/\ell(\omega))$,
assuming a diffusive regime, disregarding quantum interference effects and assuming minimum contact resistances.
The resulting frequency-dependent transmission function $\mathcal{T}(\omega)$ for different ribbon lengths is 
plotted in Fig.~\ref{fig3}, while the thermal conductance $\kappa$ using Eq.~(\ref{kappa}) is shown in 
Fig.~\ref{fig4}. The downscaling of $\mathcal{T}(\omega)$ directly impacts on $\kappa$ which is found 
to be reduced by one order of magnitude for 2~$\mu$m (at room temperature) compared with the ballistic case.
Note that the calculation time directly determines the largest $\tau$ that we can reach, the same $\tau$ corresponds to different evolution time for different frequencies, since $\tau=t/2\omega$. It takes longer calculation time for low frequency 
phonons to reach $\diff_\mathrm{max}$. Here for $\omega<70\,\mathrm{cm}^{-1}$, $\diff_\mathrm{max}$ has not been reached within the finite calculation time.
To compute the contribution of these modes to $\kappa$, a linear extrapolation for the transmission is used and the results are compared with 
those obtained from the GF method. Our analysis shows that this approximation causes an error of  less than 1.5\% for the thermal conductance at room temperature.
The conductivity of edge-disordered GNRs is obtained using $\sigma=\kappa\,L/A$, $A$ being the cross section area of the ribbons which is taken as the interplane distance of graphite layers.
At room temperature, $\sigma=1004$~Wm$^{-1}$K$^{-1}$ and $\sigma=757$~Wm$^{-1}$K$^{-1}$ with edge-disorder 10\% and 15\%, respectively for $L=$2~$\mu$m.
Comparing these values with the experimental values for suspended\cite{balandin} (3000-5000 Wm$^{-1}$K$^{-1}$) and supported\cite{RuoffScience} (600 Wm$^{-1}$K$^{-1}$) graphene, we conclude that edge-disorder is a very important source of scatterings not only in ultra-narrow ribbons, but also in GNRs as wide as at least 20 nm.
A comparison of our results with that of Ref. \onlinecite{Haldun} suggests that the effect of edge-disorder is reduced when increasing the width from 2 nm to 20 nm although it still plays an important role for large lateral sizes.
The ratio of edge atoms affected from edge reconstructions to the total number of atoms decay inversely with the width of the GNR \cite{Lan}, therefore edge profile disorder predominates over reconstruction effects with increased ribbon width.

\section{Conclusion}
{\it Conclusion}.-A new real space and order $N$ approach to compute the phonon wave packet propagation and the thermal conductance has been reported.
Diffusion coefficients and MFPs can be extracted directly from the wave packet propagation.
Its computational accuracy and efficiency were demonstrated  on disordered carbon nanotubes and large width graphene nanoribbons respectively. 
A strong impact of smooth edge-disorder on the thermal conductance was found.
Unlike edge reconstruction \cite{Lan}, edge-disorder strongly suppresses thermal conduction not only for ultra-narrow GNRs, but also for realistically large ribbons.
Phonons in edge-disordered GNRs, being scattering so effectively, pinpoints towards good thermoelectric properties of large width GNRs.
The applicability of this new method goes far beyond quasi-one dimensional systems and the area of carbon-based materials (nanotubes, graphene, etc.) studied here, and could be applied without difficulty to a wide range of other materials, including Boron-nitride-based materials \cite{savic2} or silicon-based materials (nanowires, superlattices, etc.) \cite{mads}.

{\it Acknowledgements}.-
This work was supported by the priority program Nanostructured Thermoelectrics (SPP-1386) 
of the German Research Foundation (DFG) (Contract No. CU 44/11-1), the cluster of excellence of the Free State of 
Saxony ECEMP - European Center for Emerging Materials and Processes Dresden (Project A2),
and the European Social Funds (ESF) in Saxony (research group InnovaSens), 
the ANR/P3N2009 GRAPHENE\_NANOSIM (Project No. ANR-09-NANO-016-01), and the Alexander von Humboldt Foundation.
We acknowledge support from the WCU (World Class University) program sponsored by the South Korean Ministry of Education, Science, and Technology Program, Project no. R31-2008-000-10100-0. 
The authors are thankful to N. Mingo for fruitful discussions and W.~L. thanks CAS-MPG joint doctoral program.
The Center for Information Services and High Performance Computing (ZIH) at the TU-Dresden is acknowledged.

\end{document}